\documentstyle[12pt]{article}

\skewchar\fivmi='177
\skewchar\sixmi='177
\skewchar\sevmi='177
\skewchar\egtmi='177
\skewchar\ninmi='177
\skewchar\tenmi='177
\skewchar\elvmi='177
\skewchar\twlmi='177
\skewchar\frtnmi='177
\skewchar\svtnmi='177
\skewchar\twtymi='177
\def\@magscale#1{ scaled \magstep #1}
\skewchar\fivsy='60
\skewchar\sixsy='60
\skewchar\sevsy='60
\skewchar\egtsy='60
\skewchar\ninsy='60
\skewchar\tensy='60
\skewchar\elvsy='60
\skewchar\twlsy='60
\skewchar\frtnsy='60
\skewchar\svtnsy='60
\skewchar\twtysy='60

\catcode`@=11
\def\un#1{\relax\ifmmode\@@underline#1\else
        $\@@underline{\hbox{#1}}$\relax\fi}
\catcode`@=12

\let\du=\d                      % dot-under
\def\a{\alpha}
\def\b{\beta}
\def\c{\chi}
\def\d{\delta}
\def\e{\epsilon}
\def\f{\phi}
\def\g{\gamma}
\def\h{\eta}

\def\j{\psi}
\def\k{\kappa}
\def\l{\lambda}
\def\m{\mu}
\def\n{\nu}
\def\o{\omega}
\def\p{\pi}
\def\q{\theta}
\def\r{\rho}
\def\s{\sigma}
\def\t{\tau}

\def\x{\xi}
\def\z{\zeta}
\def\D{\Delta}

\def\G{\Gamma}

\def\L{\Lambda}

\def\S{\Sigma}

\def\ve{\varepsilon}

\def\cb{{\cal B}}

\def\cd{{\cal D}}

\def\cf{{\cal F}}

\def\cm{{\cal M}}

\def\cv{{\cal V}}

\def\bo{{\raise.15ex\hbox{\large$\Box$}}}               % D'Alembertian
\def\pa{\partial}                                       % curly d
\def\de{\nabla}                                         % del
\def\pr{\prod}                                          % product
\def\TH{{\raise.2ex\hbox{$\displaystyle \bigodot$}\mskip-4.7mu \llap H \;}}
\def\face{{\raise.2ex\hbox{$\displaystyle \bigodot$}\mskip-2.2mu \llap {$\ddot
        \smile$}}}                                      % happy face
\def\dg{\sp\dagger}                                     % hermitian conjugate
\def\sp#1{{}^{#1}}                              % superscript (unaligned)
\def\Tilde#1{\widetilde{#1}}                    % big tilde
\def\Hat#1{\widehat{#1}}                        % big hat
\def\ket#1{\left| #1\right\rangle}              % | >
\def\VEV#1{\left\langle #1\right\rangle}        % < >
\def\abs#1{\left| #1\right|}                    % | |
\def\leftrightarrowfill{$\mathsurround=0pt \mathord\leftarrow \mkern-6mu
        \cleaders\hbox{$\mkern-2mu \mathord- \mkern-2mu$}\hfill
        \mkern-6mu \mathord\rightarrow$}
\def\dvec#1{\vbox{\ialign{##\crcr
        \leftrightarrowfill\crcr\noalign{\kern-1pt\nointerlineskip}
        $\hfil\displaystyle{#1}\hfil$\crcr}}}           % <--> accent
\def\frac#1#2{{\textstyle{#1\over\vphantom2\smash{\raise.20ex
        \hbox{$\scriptstyle{#2}$}}}}}                   % fraction
\def\ha{\frac12}                                        % 1/2
\def\sfrac#1#2{{\vphantom1\smash{\lower.5ex\hbox{\small$#1$}}\over
        \vphantom1\smash{\raise.4ex\hbox{\small$#2$}}}} % alternate fraction
\def\bfrac#1#2{{\vphantom1\smash{\lower.5ex\hbox{$#1$}}\over
        \vphantom1\smash{\raise.3ex\hbox{$#2$}}}}       % "
\def\afrac#1#2{{\vphantom1\smash{\lower.5ex\hbox{$#1$}}\over#2}}    % "
\def\[{\lfloor{\hskip 0.35pt}\!\!\!\lceil}
\def\]{\rfloor{\hskip 0.35pt}\!\!\!\rceil}
\def\Lag{{\cal L}}
\def\du#1#2{_{#1}{}^{#2}}
\def\ud#1#2{^{#1}{}_{#2}}

\def\fracm#1#2{\hbox{\large{${\frac{{#1}}{{#2}}}$}}}
\def\half{{\fracm12}}
\def\ha{\half}
\def\tr{{\rm tr}}

\def\ula{{\underline a}} \def\ulb{{\underline b}} \def\ulc{{\underline c}}
\def\uld{{\underline d}}

\def\un{\underline}
\def\fracmm#1#2{{{#1}\over{#2}}}

\def\low#1{{\raise -3pt\hbox{${\hskip 0.75pt}\!_{#1}$}}}

\def\Dot#1{\buildrel{_{_{\hskip 0.01in}\bullet}}\over{#1}}

\def\Tilde#1{{\widetilde{#1}}\hskip 0.015in}
\def\Hat#1{\widehat{#1}}

\newskip\humongous \humongous=0pt plus 1000pt minus 1000pt
\def\caja{\mathsurround=0pt}
\def\eqalign#1{\,\vcenter{\openup2\jot \caja
        \ialign{\strut \hfil$\displaystyle{##}$&$
        \displaystyle{{}##}$\hfil\crcr#1\crcr}}\,}
\newif\ifdtup

\def\ref#1{$\sp{#1)}$}

\def\pl#1#2#3{Phys.~Lett.~{\bf {#1}B} (19{#2}) #3}
\def\np#1#2#3{Nucl.~Phys.~{\bf B{#1}} (19{#2}) #3}
\def\prl#1#2#3{Phys.~Rev.~Lett.~{\bf #1} (19{#2}) #3}
\def\pr#1#2#3{Phys.~Rev.~{\bf D{#1}} (19{#2}) #3}

\def\jmp#1#2#3{J.~Math.~Phys.~{\bf {#1}} (19{#2}) #3}
\def\ap#1#2#3{Ann.~of Phys.~{\bf {#1}} (19{#2}) #3}
\def\prep#1#2#3{Phys.~Rep.~{\bf {#1}C} (19{#2}) #3}

\def\mpl#1#2#3{Mod.~Phys.~Lett.~{\bf A{#1}} (19{#2}) #3}

\def\ibid#1#2#3{{\it ibid.}~{\bf {#1}} (19{#2}) #3}

\parskip=\medskipamount                 % space between paragraphs (LaTeX)
\thicklines                         % thick straight lines for pictures (LaTeX)

\begin{document}

\thispagestyle{empty}               % no heading or foot on title page (LaTeX)

\def\border{                                            % border
        \setlength{\unitlength}{1mm}
        \newcount\xco
        \newcount\yco
        \xco=-24
        \yco=12
        \begin{picture}(140,0)
        \put(-20,11){\tiny Institut f\"ur Theoretische Physik Universit\"at
Hannover~~ Institut f\"ur Theoretische Physik Universit\"at Hannover~~
 Institut f\"ur Theoretische Physik Hannover}
        \put(-20,-241.5){\tiny Institut f\"ur Theoretische Physik Universit\"at
Hannover~~ Institut f\"ur Theoretische Physik Universit\"at Hannover~~
 Institut f\"ur Theoretische Physik Hannover}
        \end{picture}
        \par\vskip-8mm}

\def\headpic{                                           % UH heading
        \indent
        \setlength{\unitlength}{.8mm}
        \thinlines
        \par
        \begin{picture}(29,16)
        \put(75,16){\line(1,0){4}}
        \put(80,16){\line(1,0){4}}
        \put(85,16){\line(1,0){4}}
        \put(92,16){\line(1,0){4}}

        \put(85,0){\line(1,0){4}}
        \put(89,8){\line(1,0){3}}
        \put(92,0){\line(1,0){4}}

        \put(85,0){\line(0,1){16}}
        \put(96,0){\line(0,1){16}}
        \put(79,0){\line(0,1){16}}
        \put(80,0){\line(0,1){16}}
        \put(89,0){\line(0,1){16}}
        \put(92,0){\line(0,1){16}}

        \put(79,16){\oval(8,32)[bl]}
        \put(80,16){\oval(8,32)[br]}

        \end{picture}
        \par\vskip-6.5mm
        \thicklines}

\border\headpic {\hbox to\hsize{
ITP--UH--1/93 \hfill February 1993}}\par
\vskip2cm
\begin{center}
{\Large\bf Space-Time Supersymmetry of Extended Fermionic Strings in
$2 + 2$ Dimensions} \\
\vglue.2in
%{\it Intern. Physics Classification Nos.} 1117, 1130, 0465\\
\vglue.2in
Sergei V. Ketov\footnote{On leave of absence from: High
Current Electronics Institute of the Russian Academy of Sciences, Siberian
Branch, Akademichesky~4, Tomsk 634055, Russia} \\
\vglue.2in
{\it Institut f\"ur Theoretische Physik, Universit\"at Hannover}\\
{\it Appelstra\ss{}e 2, W-3000 Hannover 1, Germany}
\end{center}

\begin{center}
{\Large\bf Abstract}
\end{center}

The $N=2$ fermionic string theory is revisited in light of its recently
proposed equivalence to the non-compact $N=4$ fermionic string model. The
issues of space-time Lorentz covariance and supersymmetry for the BRST
quantized $N=2$ strings living in uncompactified $2 + 2$ dimensions are
discussed. The equivalent local quantum supersymmetric field theory appears
to be the most transparent way to represent the space-time symmetries of the
extended fermionic strings and their interactions. Our considerations support
the Siegel's ideas about the presence of $SO(2,2)$ Lorentz symmetry as well
as at least one self-dual space-time supersymmetry in the theory of the
$N=2(4)$
fermionic strings, though we do not have a compelling reason to argue about
the necessity of the {\it maximal} space-time supersymmetry. The world-sheet
arguments about the absence of all string massive modes in the physical
spectrum, and the vanishing of all string-loop amplitudes in the Polyakov
approach, are given on the basis of general consistency of the theory.

\newpage

\section{Introduction}

The theory of the $N=2$ fermionic strings is a quite natural extension of the
conventional NSR string theory: the latter is based on the {\it gauged}
$N=1$ superconformal symmetry of the string world-sheet, while the former uses
the $N=2$ extended local superconformal symmetry (see ref.~\cite{nm} for a
recent review). The string theory methods always look like a bridge between
the two-dimensional (world-sheet) and higher-dimensional (space-time) concepts,
 as well as  the corresponding symmetries. One of the famous relations
of that type is the correspondence between the $N=2$ {\it global}
superconformal symmetry on the superstring world-sheet and the $N=1$
space-time sypersymmetry in the effective four-dimensional field theory
resulting from the superstring compactification. Since the
four-dimensional space-time supersymmetry can naturally be extended up to the
$N=4$ (if the maximal spin is 1) or up to the $N=8$ (if the maximal spin
is 2), one may ask about the existence of the {\it critical} string models
leading to the $N=4$ or $N=8$ supersymmetric effective four-dimensional field
theories {\it without} any compactification. The similar motivation has been
used in the past to argue about the relevance of supermembranes, which,
contrary to superstrings, have a natural room for the 11-dimensional (maximally
extended!) supergravity \cite{bst}.

The $N=2$ fermionic strings, being a natural extension of the conventional
superstrings, have a rather controversal status. They were first introduced
and investigated in two real space-time dimensions, their space-time
symmetries, if any, were
always quite obscure, and the understanding of the $N=2$ string amplitudes is
still lacking. Recently, some new important developments have taken place, and
 they are going to change the whole status of the theory. In this paper we
review some of the recent new results about the structure of the extended
fermionic strings, when a special attention being paid on the status of
space-time supersymmetry and its adequate description in that theories. The
relevant references are cited in parallel with the discussion.

The motivation to study the $N=2$ and $N=4$ fermionic strings is at least
two-fold.
On the one side, they are just the useful polygons for analyzing the specific
properties of the more complicated conventional superstrings. On the other
hand,
being intimately related to the self-dual field theories and, hence, to the
integrable models \cite{wd}, the extended fermionic strings could be used for
the quantization of the latter.

Another motivation is just to relate the maximal gauged superconformal
symmetry
on the world-sheet with the maximal (conformal) sypersymmetry in space-time.

\newpage

\section{World-Sheet Symmetries and Actions}

The $N=2$ superconformal algebra (SCA) in two dimensions\footnote{The
Euclidean two dimensions are supposed to form a Riemann surface $\S$
parametrized by a complex variable $z$ in local coordinate charts.} comprises
a stress tensor $T(z)$ of (conformal) dimension $2$, two real supercharges
$G^i(z)$ of dimension $3/2$, and an Abelian current $J(z)$ of dimension $1$,
with the OPE \cite{a1,a2,a3,zf}
$$T(z)T(w)\sim \fracmm{c}{2(z-w)^4}+\fracmm{2}{(z-w)^2}T(w)+\fracmm{1}{z-w}\pa
T(w)~,$$
$$T(z)G^i(w)\sim \fracmm{3G^i(w)}{2(z-w)^2}+\fracmm{1}{z-w}\pa G^i(w)~,$$
$$G^i(z)G^j(w)\sim \left[\fracmm{2c}{3(z-w)^3}+\fracmm{2T(w)}{z-w}\right]
\d^{ij}+i\left[\fracmm{2J(w)}{(z-w)^2}+\fracmm{1}{z-w}\pa J(w)\right]\ve^{ij}~,
$$
$$T(z)J(w)\sim \fracmm{J(w)}{(z-w)^2}+\fracmm{1}{z-w}\pa J(w)~,$$
$$J(z)G^i(w)\sim i\ve^{ij}\fracmm{G^j(w)}{z-w}~,$$
$$J(z)J(w)\sim \fracmm{c}{3(z-w)^2}~,\eqno(2.1)$$
where the central charge $c$ has been introduced, $i=1,2$. The internal
symmetry group corresponding to the Abelian Kac-Moody (KM) current $J(z)$ can
be either {\it compact} $U(1)$, or {\it non-compact} $GL(1)$.

The Abelian internal symmetry of the $N=2$ super-Virasoro algebra (2.1) can be
extended to a non-Abelian symmetry in the $N=4$ SCA \cite{a1,a3}, without
introducing new ``subcanonical'' charges.\footnote{By ``subcanonical'' charges
\cite{a1} one means charges of dimension other than $2$, $3/2$ or $1$.
The canonical charges are just those which correspond to conformal symmetry,
supersymmetry and internal symmetry, respectively.} The $N=4$ SCA has a {\it
complex doublet}
of supersymmetry charges, $G_{\a}(z)$ and $\bar{G}_{\a}(z)$, transforming in
the fundamental representation of the internal symmetry group $G$, which can
also be either {\it compact} $SU(2)$ or {\it non-compact} $SU(1,1)$.\footnote{
In fact, only the compact cases were the subjects of investigation in the
early studies of the corresponding dual models \cite{a1,a2,a3}.} The internal
symmetry generators $J^i(z)$ of the $N=4$ SCA form the KM algebra $\Hat{G}=
\Hat{SU(2)}$, or $\Hat{G}=\Hat{SU(1,1)}$, respectively, the KM level being
fixed by the central charge $c$:
$$J^I(z)J^J(w)\sim \fracmm{if^{IJK}}{z-w}J^K(w)+\fracmm{c}{3(z-w)^2}\d^{IJ}~,
\eqno(2.2)$$
where the structure constants $f^{IJK}$ of the internal symmetry group $G$ have
been introduced, $I,J,K=1,2,3$. Given the  transformation properties of the
$N=4$ SCA charges, the rest of the $N=4$ SCA OPE's is quite obvious.

All the $N>2$ SCA's are known to contain subcanonical charges needed to close
an algebra \cite{a1,a3}, with just the two exceptions at $N=4$, because of the
relevant group decompositions
$$SO(4)\cong SU(2)\otimes SU(2)~,$$
$$SO(2,2)\cong SU(1,1)\otimes SU(1,1)~.\eqno(2.3)$$
It is eq.~(2.3) that allows to restrict the internal symmetry generators of the
$N=4$ SCA to be {\it self-dual} (SD) or {\it anti-self-dual} (ASD), and thus
close the algebra, when using the minimal set of canonical charges only.

Being realized globally, both $N=2$ and $N=4$ SCA's arise in describing
compactification of the conventional ten-dimensional superstrings down to four
dimensions, when space-time supersymmetry of the effective four-dimensional
theory is required \cite{cgsw,g8}. The unitary representations and their
characters have been constructed in the past for $N=2$ \cite{bfk,g8}, as well
as for $N=4$ \cite{et}.

The $N$-extended fermionic string models \cite{a1,a2,a3} arise when the
$N$-extended SCA's are {\it locally} realized, i.e. gauged.  The bridge between
strings and CSA's is provided by free field realizations of the latter. In the
absence of subcanonical charges in a SCA, there always exists its free field
realization in terms of canonical fields to be interpreted as bosonic and
fermionic coordinates \cite{a1,a2,a3}.

The convenient fermionic string (free field) realization of the $N$-extended
SCA is provided by the world-sheet covariant gauge-fixed form of the
supercurrent
multiplet in the (locally) $N$-supersymmetrized Polyakov string action
\cite{p}. It can be constructed by coupling the $N$-extended supergravity to
the $N$-extended scalar multiplet in two dimensions. In the case of $N=2$, the
 appropriate action (in components) has been constructed by Brink and Schwarz
\cite{bs1}, and it takes the form
$$I_{\rm BS} = \fracmm{1}{\p}\int d^2z\,e\left\{ \fracm{1}{2}g^{pq}\pa_p
x^{*}_{\m}\pa_q x^{\m}+\fracm{i}{2}\bar{\j}_{\m}\g^p\dvec{D}_p\j^{\m}\right.$$
$$\left. +A_p\left(\bar{\j}_{\m}\g^p\j^{\m}\right)+\left[\left(\pa_p x^{\m} +
\bar{\c}_p\j^{\m}\right)\left(\bar{\j}_{\m}\g^q\g^p\c_q\right)+{\rm h.c.}
\right]\right\}~,\eqno(2.4)$$
where a set of $D$ {\it complex} scalar (``matter'') multiplets $(x,\j)$ is
coupled to the $N=2$ supergravity multiplet $(e_p^{\tilde{p}},\c_p,A_p)$ in
two dimensions; $p,q=1,2$; $\m=1,2$. The spinors $\j$ and $\c$ are both
complex,
whereas the bosonic zweibein $e_p^{\tilde{p}}$ and the Abelian vector gauge
field $A_p$ are real. The $D_p$ represents the standard gravitational covariant
derivative for spinors \cite{n81,pr87}. The local world-sheet gauge symmetries
of the BS-action (2.4) are \cite{bs1,ft1}:
$$\eqalign{
(i)~ &~~ reparametrization~ invariance~,\cr
(ii)~ &~~ Lorentz~ invariance~,\cr
(iii)~ &~~ N-extended~supersymmetry~,\cr
(iv)~ &~~ scale~invariance~,\cr
(v)~ &~~ N-extended~conformal~supersymmetry~,\cr
(vi)~ &~~ phase~ and~chiral~gauge~invariance~.\cr}\eqno(2.5)$$
The symmetries (i), (ii), (iv) and (vi) are bosonic, the symmetries (iii) and
(v) are fermionic.

Introducing the oscillator representation for the real bosonic coordinates
$(x^{\m}=x_1^{\m}+ix_2^{\m})$ and the real fermionic coordinates $(\j^{\m}=
\j_1^{\m}+i\j_2^{\m})$ in terms of their real modes $\a_{in}^{\m}$ and
$d_{ir}^{\m}$ respectively, the Gupta-Bleuler quantization implies, as usual
in string theory, the relations \cite{gsw,a2}
$$\eqalign{
\[ \a^{\m}_{in},\a^{\n}_{jm} \] = & \d_{ij}\h^{\m\n}n\d_{n+m,0}~,\cr
\{ d^{\m}_{ir},d^{\n}_{js} \} = & \d_{ij}\h^{\m\n}n\d_{r+s,0}~,\cr}\eqno(2.6)$$
where the modding for spinors can be either integer $(r,s\in{\bf Z})$ or
half-integer
$(r,s\in{\bf Z}+1/2)$, depending on boundary conditions.

The local gauge symmetries (2.5) can then be used to gauge away all the
supergravity
fields, leaving the dependence on their moduli only. The vanishing
supercurrent
multiplet of the BS theory (2.4) determines the constraint algebra, which
is just the
realization of the $N=2$ SCA (2.1) in terms of the free oscillators (2.6):
$$T(z) = \fracmm{1}{2}:P_i(z)\cdot P_i(z):-\fracmm{i}{2}:\j_i(z)\cdot
\pa\j_i(z):~,$$
$$G_1(z)=\j_i(z)\cdot P_i(z)~,\qquad G_2(z)=\ve_{ij}\j_i(z)\cdot P_j(z)~,$$
$$J(z)=\fracmm{i}{4}\ve_{ij}\j_i(z)\cdot \j_j(z)~,\eqno(2.7)$$
where the fields\footnote{When dealing with chiral (say, left-moving)
modes of a
closed string, it is just enough to consider the holomorphic dependence
on $z$.}
$$P^{\m}_i(z)= i\pa x^{\m}_i(z) =\sum_n \a^{\m}_{in}z^{-n}~,$$
$$\j^{\m}_i(z)=\sum_s d^{\m}_{is}z^{-s}~,\eqno(2.8)$$
have been introduced. The dots in eq.~(2.7) mean contractions of target
space vector
indices with respect to a flat orthogonal metric $\h^{\m\n}$ to be
determined below. The
$N=2$ fermionic string theory apparently has the $SO(2)\otimes SO(D)$ or
$SO(1,1)\otimes SO(D-q,q)$, $q=0,\ldots,D-1$, as the ``Lorentz'' group.

The $N=4$ counterpart to the BS action (2.4) has been constructed by
Pernici and
Nieuwenhuizen \cite{pn8}. Their $N=4$ fermionic string action is in fact quite
similar to that of eq.~(2.4), and takes the form
$$I_{\rm PN} = \fracmm{1}{\p}\int d^2z\,e\left\{ \fracm{1}{2}g^{pq}\pa_p
x^{*}_{\m}\pa_q x^{\m}+\fracm{i}{2}\bar{\j}_{\m\m'}\g^p\dvec{D}_p\j^{\m\m'}
\right.$$
$$\left. +A_p^I\left(\bar{\j}_{\m\m'}\g^p\s_I^{\m'\n'}\j^{\m\n'}\right)+
\left[\left(
\pa_p x^{\m} +\bar{\c}_{p\m'}\j^{\m\m'}\right)\left(\bar{\j}_{\m\n'}\g^q\g^p
\c_q^{\n'}\right)+{\rm h.c.}\right]\right\}~,\eqno(2.9)$$
where the fields $x$, $\j$ and $\c_p$ are now considered to be
{\it quaternionic}.\footnote{It means that the second independent complex
structure
(``$j$'' or ``prime'') has been introduced, so that $x=x_1 + jx_2,~\j^{\m'}
=\j_1 +
j\j_2$, and $x_1=x_{11} + ix_{12},~x_2=x_{21}-ix_{22}$, and similarly for
fermions.
The quaternionic index can equally be represented as a multi-index $(\a\m')$,
where
$\a=1,2$ and $\m'=1',2'$.} The $\s_I$ are the $SU(2)$ or $SU(1,1)$ generators
in the fundamental representation.  The $N=4$ analogues to eqs.~(2.5), (2.6),
(2.7) and
 (2.8) are now quite obvious. In particular, one gets \cite{a3}
$$T(z) = \fracmm{1}{2}:P_{\a\m'}(z)\cdot P^{\a\m'}(z):-\fracmm{i}{2}:
\j_{\m'}(z)\cdot
\pa\j^{\m'}(z):~,$$
$$G_{\a}(z)=\j^{\m'}(z)\cdot P_{\a\m'}(z)~,$$
$$J^I(z)=\fracmm{i}{2}\j^{\m'}(z)\cdot\s_{\m'\n'}^I\j^{\n'}(z)~,
\eqno(2.10)$$
where all the fermions have been chosen to be the complex doublets with
 respect to
$SU(2)$
or $SU(1,1)$. Notably, in eqs.~(2.9) and (2.10) the index $\m$ can be
combined with the
index $\m'$ into one vector index so that, when being compared to the BS
theory, the
PN theory (2.9) is actually invariant under a larger ``Lorentz'' group
$SO(2D-2q,2q)$,
{\it provided} there is at least one factor of $SU(2)$ or $SU(1,1)$ in the
decomposition
of the $SO(2D-2q,2q)$ into simple factors.

Therefore, there are clearly {\it two} distinct extended fermionic string
models both
for the $N=2$ and $N=4$: one with the compact internal symmetry group, and
another one
with the non-compact group.

\section{BRST and BFV}

The Becchi-Rouet-Stora-Tyutin (BRST) quantization prescription
\cite{brs,t1} implies
certain conditions on an initial classical constrained system to satisfy.
In light of
the more general Batalin-Fradkin-Vilkovisky (BFV) quantization prescription
\cite{fv,bv1,bv2}, the constraints are to be (1) of the {\it first class},
and (2) {\it
 irreducible}.

The standard Dirac theorem \cite{d} explains the way how to quantize a
constrained
system, first introducing the independent (physical) canonically conjugated
variables
arising from the solution of the constraints subject to admissible gauges,
then
quantizing that physical variables canonically, and finally rewriting the
result back to
the initial phase space. The Dirac's quantization is obviously consistent
with the
canonical one, unitary but non-covariant. The BRST prescription is just the
Dirac's
quantization in the covariant form, when a covariant (with respect to the
 Lorentz
transformations) gauge is chosen. The covariance is maintained in the BRST
quantization by extending the initial phase space by ghosts, so that the
extended
(fields + ghosts) system can be quantized ``naively'', when using the
BRST-invariant
Hamiltonian and constraints, by integrating over the ghosts in the
quantum generating
functional. To be specific, given bosonic (B) and fermionic (F) constraints
satisfying
a {\it closed} algebra\footnote{The indices used below in eqs.~(3.1)--(3.5)
have the
meaning different from that used in the bulk of the paper, however they
could hardly
be confused.}
$$\[B^a,B^b\]=f\ud{ab}{c}B^c~,\quad \[B^a,F^{\b}\]=f\ud{a\b}{\g}F^{\g}~,\quad
\{F^{\a},F^{\b} \}=f\ud{\a\b}{c}B^c~,\eqno(3.1)$$
where $B^0\equiv H_0$ is an initial Hamiltonian, and $f$'s are the constraint
algebra
structure {\it constants}, the BRST-invariant quantities are defined by
\cite{bv1}
$$\cb^a=\{ \r^a,Q\}~,\qquad \cf^{\a} =\[ \x^{\a},Q \]~.\eqno(3.2)$$
In eq.~(3.2) the canonically conjugated  ghosts for each constraint,
$$ {\rm B}:~~ \r^a,\h_b~,\qquad {\rm F}:~~ \x^{\a},\l_{\b}~,\eqno(3.3)$$
have been introduced. By definition, they have statistics opposite to that
of the
constraints and satisfy the (anti)commutation relations:
$$\{ \r^a,\h_b \} = \d_{ab}~,\qquad \[ \x^{\a}, \l_{\b} \] = \d_{\a\b}~.
\eqno(3.4)$$
The operator $Q$ introduced in eq.~(3.2) is known as the BRST charge
\cite{bv1}:
$$Q=B^a\h_a +F^{\a}\l_{\a}-\fracm{1}{2}f\ud{ab}{c}\r^c\h_a\h_b-
f\ud{a\b}{\g}\x^{\g}\h_a
\l_{\b}-\fracm{1}{2}f\ud{\a\b}{c}\r^c\l_{\a}\l_{\b}~.\eqno(3.5)$$
Classically, one has $\{Q,Q\}_{\rm PB}=0$ with respect to the Poisson
bracket. Quantum
mechanically, the consistent quantization requires the BRST operator
to be nilpotent,
$Q^2=0$.

Given the {\it reducible} constraints, the BRST rules have to be modified
\cite{bv2}. The BFV-prescription introduces the new ghosts beyond those
needed in
the BRST framework, when dealing with the reducilbe constraints. The
derivation of the
generalized BFV-BRST rules also goes back and forth: first one chooses
an irreducible
subset of constraints, applies BRST rules, and then rewrites the quantized
theory in
terms of the initial variables and constraints by using ``ghosts for
ghosts'' \cite{bv2}.

The BRST techniques have been applied to the quantization of the $N=2$
fermionic
string in refs.~\cite{bi,mm1,jb1} (see ref.~\cite{mm1}, as for the BRST
quantization
of the $SU(2)$ fermionic string of Ademollo et al \cite{a3}). It is easy
to check the
irreducibility of the $N=2$ first-class constraints (2.7), which justifies
the
applicability of the BRST rules to that case. In the oscillator
representation, the BRST
ghosts are
$$T(z)\to~T_n\to~\r_n,\h_m~:~~~\{ \r_n,\h_m \} = \d_{n+m,0}~,$$
$$G^i(z)\to~G_r^i\to~\x^i_r,\l^j_s~:~~~\[ \x^i_r,\l^j_s \] =\d^{ij}\d_{r+s,0}~,
\eqno(3.6)$$
$$J(z)\to~J_n\to~\o_n,\f_m~:~~~\{ \o_n,\f_m \} = \d_{n+m,0}~,$$
in accordance to eqs.~(3.3) and (3.4). The BRST charge reads \cite{bi}:
$$Q = \fracmm{1}{2\p i}\oint_0 \fracmm{dz}{z}\,:\left\{ T(z)\h(z) +
G^i(z)\l_i(z)+
J(z)\f(z) + z\r(z)\h(z)\h'(z)\right.$$
$$-\fracm{1}{2}z\x_i(z)\h'(z)\l_i(z) + z\x_i(z)\h(z)\l'(z) -
\r(z)\l_i(z)\l_i(z)
+z\o(z)\h(z)\f'(z)$$
$$\left. -\fracm{i}{2}\ve^{ij}\x_j(z)\f(z)\l_i(z)+
2i\ve^{ij}z\o(z)\l_i(z){\l'}_j(z)
-\a_0\h(z)\right\}:~,\eqno(3.7)$$
where the ghost fields $\h(z)=\sum_n\h_nz^{-n}$ etc, and the ``intercept''
 (normal
ordering ambiguity) constant $\a_0$ have been introduced. It is now
straightforward
to calculate that {\cite{bi}
$$Q^2 = \fracmm{1}{4\p i}\oint_0 dz\,:\left\{ \fracm{1}{4}(D-2)(z\h)'''z\h +
(D-2)z\l_i''\l_i + \fracm{1}{4}(D-2)\f'\f\right.$$
$$\left. + 2\a_0\h'\h+\left[\fracm{1}{4}(2-D)+2\a_0\right]\l_i\l_i/z\right\}:~,
\eqno(3.8)$$
where the half-integer modding for the matter spinors $\j$'s has been used.
 For the
integer modding, the result is quite similar to that of eq.~(3.8): one has
only to
exchange the numerical coefficients between the last two terms in the curved
brackets
\cite{bi}. Hence, the critical $N=2$ fermionic strings have \cite{ft1,bi,mm1}
$$ D=2~,\qquad \a_0 = 0~.\eqno(3.9)$$
This critical dimension also follows from the conformal anomaly counting
\cite{ft1}:
$$\begin{array}{ccccccc}
(\r,\h) & {} & (\x^i,\l^i) & {} & (\f,\o) & {} &   (x^i ,\j^i)^{\m}    \\
  -26   &  + &  2\cdot 11  &  + &   -2    &  + & 2\cdot D\cdot(1 +
\fracm{1}{2})~,
\end{array}\eqno(3.10)$$
whose vanishing also yields $D=2$.

The fact that the $N=2$ fermionic strings live in $2$ {\it complex}
dimensions has
been known for a long time \cite{a2,ft1}, however it was not known until
recently
how to introduce interactions in $2$ complex or $4$ real dimensions.
That's why the
early studies of the $N=2$ fermionic string amplitudes were confined to the
$2$
{\it real} dimensions \cite{a2,gr}. The choice of two real dimensions was
also
suggested by the apparently $2$-dimensional ``Lorentz'' or ``space-time''
group
$SO(2)\otimes SO(2)$ or $SO(1,1)\otimes SO(1,1)$. The four-dimensional
space-time
interpretation of the $N=2$ fermionic string theory was first suggested
by D'Adda and
Lizzi \cite{al}, who showed how to rewrite its defining constraints in
the $SO(2,2)$
covariant way.

The strange ``hidden'' four-dimensional Lorentz symmetry of the $N=2$
theory (2.4) can
be understood, when turning to the quantization of the $N=4$ fermionic
string theory
(2.9). The analysis of the $N=4$ constraints (2.10) reveals an unexpected
result:
they are irreducible in the compact $SU(2)$ case, but become {\it reducible}
in the
non-compact $SU(1,1)$ case, as has been noticed recently by Siegel
\cite{s1}. Being
a subset of the $N=4$ SCA constraints (2.10), the $N=2$ SCA constraints
(2.7) already
eliminate all the excited string states but the ground states, so that
the rest of the
$N=4$ constraints in eq.~(2.10) becomes redundant. The BRST analysis of
the spectrum
of the $N=2$ theory shows that all the physical states are the highest
weight states
of the $N=4$ SCA, the other being either non-physical or pure gauge
with respect to
the $N=2$ SCA \cite{jb1}. Therefore, the $N=2$ non-compact fermionic
string theory and
the $N=4$ non-compact fermionic string theory are in fact equivalent,
the latter being
the covariant form of the former with respect to the ``Lorentz''
group $SO(2,2)$. In
its turn, the non-compact $N=2$ theory can be interpreted as a
``partially gauged''
non-compact $N=4$ theory \cite{s1}. The absence of the massive string
modes in this
theory also follows from a calculation of the partition
function\footnote{For
definiteness, the case of closed $N=2$ strings with anti-periodic
(for fermions)
boundary conditions has been chosen.} \cite{bh,ov1}:
$$Z_{\rm AAAA}={\rm sTr}\left[ q^{T_0}\bar{q}^{\bar{T}_0}t_{\rm L}^{J_0}
t_{\rm R}^{\bar{J}_0}\right]=\int d^4p\,(q\bar{q})^{p^2/2}~,\eqno(3.11)$$
where $q=\exp(2\p i\t)$, $t_{\rm L,R}=\exp(2\p i\q_{\rm L,R})$. Stated
differently,
the non-zero modes of $(x,\j)$ cancel with those of ghosts.

The very general relation between the mass $m_0$ of a ground state and a
critical
dimension $D$,
$$\a' m^2_0 = -\fracmm{1}{24}(D-2)\left( \#_{\rm B} + \fracm{1}{2}\#_{\rm F}
\right)~,
\eqno(3.12)$$
derived by Brink and Nielsen \cite{bn8} from their analysis about zero-point
fluctuations in dual string models, forces the ground state in the $N=2$
string model
to be always massless, $p^2=0$. In particular, the ground state of the
Euclidean $SO(4)$
 theory is therefore an identity with $p=0$ and no dynamics, unless complex
values of
momenta are allowed, but then it would be just the ``doubly Wick-rotated''
$SO(2,2)$
theory. The necessity of a ``complex'' time in the $N=2$ fermionic string
theory was
also argued by Ooguri and Vafa \cite{ov1} from various viewpoints.

The $N=4$ fermionic string with the {\it compact} internal symmetry group
\cite{a3}
still has the irreducible constraints (2.10). Hence, it can be quantized
along the
standard BRST lines, which yield the negative critical dimension $D=-2$
\cite{mm1}.
It has been known for a long time that this theory suffers from inevitable
ghosts both
in its spectrum \cite{a3}, and in its amplitudes \cite{bfy}. This conclusion
also
follows from the conformal anomaly counting in the $SU(2)$ fermionic string:
$$\begin{array}{ccccccc}
(\r,\h) & {} & (\x_i^a,\l^i_a) & {} & (\f^I,\o_I) & {} &   (x_{i\m'} ,
\j_{i\m'})^{\m} \\
  -26   &  + & 2\cdot 2\cdot 11  &  + & -3\cdot 2 & + & 2\cdot 2\cdot D
\cdot(1 +
\fracm{1}{2})~,\end{array}\eqno(3.13)$$
whose cancellation implies $D=-2$.

The reducibility of the $N=4$ constraints means a linear dependence between the
generators of the gauged $SU(1,1)$ superconformal algebra. It can be understood
in part by considering the gauge transformation laws of the gravitino and the
$SU(1,1)$ vector fields in the PN theory, which are themselves invariant
under the
Abelian complex transformations of the parameters of superconformal and
internal
symmetry. The $N=4$ harmonic
superspace with harmonic coordinates parametrizing the coset $SU(1,1)/GL(1)$
should therefore be quite appropriate for a covariant description of the $N=4$
theory, and it makes indeed the gauge symmetry generator dependence to be
explicit \cite{s1}. The BFV quantization implies the appearance of the
``ghosts for ghosts'' according to the rule \cite{bv1,bv2}:
$$\begin{array}{ccccc}
&&G^a&&\\
&\x^a&&\l_a&\\
\e^{(1)}&&\x^{(1)}&&\l_{(1)} \\
\cdots&\cdots&&\cdots&\cdots\\
\end{array}\qquad
\begin{array}{ccccc}
&&J^I&&\\
&\f^I&&\o_I&\\
\z^{(1)}&&\f^{(1)}&&\o_{(1)} \\
\cdots&\cdots&&\cdots&\cdots\\
\end{array}\eqno(3.14)$$
The {\it complex} Abelian fermionic and bosonic ghosts,
$(\x^{(n)}_i,\l^i_{(n)})$
and  $(\f_i^{(n)},\o^i_{(n)})$ respectively, contribute to the conformal
anomaly.
Therefore, the conformal anomaly counting (3.13) gets to be modified,
presumably as
\footnote{The way of treating here the anomalies of the ``ghosts for ghosts''
is
quite similar to that used in the BFV quantization of the Green-Schwarz
superstring
\cite{kr}. In particular, one uses $1-1+1-1+\ldots =1/2$.}
$$\begin{array}{ccccccc}
(\r,\h) & {} & (\x_i^a,\l^i_a) & {} & (\f^I,\o_I) & {} &   (x_{i\m'} ,
\j_{i\m'})^{\m} \\
-26 & + & 2\cdot 2\cdot 11&  + & -3\cdot 2 & + & 2\cdot 2\cdot D\cdot(1+
\fracm{1}{2})\\
    &   &(\x_i^{(1)},\l^i_{(1)})& {} & (\f_i^{(1)},\o^i_{(1)}) & {} &   \\
    & + & -2\cdot 11 & + & 2\cdot 2 & &  {}
\end{array}\eqno(3.15)$$
Its cancellation now implies one quaternionic $(D=1)$  or four real dimensions.

Since the critical non-compact $N=2$ (or $N=4$) fermionic string implies the
four-dimensional
target (``space-time'') of the signature $(+,+,-,-)$ and the ``Lorentz''
group $SO(2,2)$
for its consistent propagation, and it does not have any massive modes in
its spectrum,
this string theory should be equivalent to a local four-dimensional quantum
field
theory, which would describe the $N=2(4)$ string ground states and their
interactions.
The spectrum of the equivalent effective field theory and its vertices have
to reproduce
the $N=2$ string $S$-matrix. In order to identify the effective field theory
of the $N=2$
fermionic strings in $2 + 2$ space-time dimensions, the $N=2$ string ground
states and their scattering amplitudes should therefore be considered. Being
restricted to the on-shell quantities, the $N=2$ (non-covariant) string
formalism is clearly more convenient for those purposes than the $N=4$
(covariant) one. We are thus going to discuss first the non-covariant
constraints on
the effective field theory, and then propose its covariant (in the
$(2+2)$-dimensional
``space-time'') action to be fixed by the symmetries and spectrum of the
$N=2(4)$
fermionic string theory.

\section{Spectrum and Tree Scattering Amplitudes}

We now analyze the spectrum of the physical states in the $N=2(4)$ fermionic
string theory. To be specific, let's consider the left-moving (analytic) modes
of the closed string propagating in $2 + 2$ space-time dimensions.

The physical states are all the ground states in the $N=2$ theory, which
can equally be represented in an $SO(2,2)$ covariant way as the ground states
of the $N=4$ theory.  All four real world-sheet fermions should then be
considered
on equal footing, when their boundary conditions are analyzed. A real
world-sheet
fermion $\j(z)$ can be either {\it periodic} ($P$) or {\it anti-periodic}
($A$):
$$\j^{\ula}(2\p)=\pm\j^{\ula}(0)~,\qquad \ula = (a,\bar{a}) =1,2,3,4~,
\eqno(4.1)$$
which implies 16 different sectors in the theory:
$$AAAA$$
$$PAAA\quad APAA \quad AAPA \quad AAAP $$
$$ PPAA \quad PAPA \quad PAAP \quad APPA \quad APAP \quad AAPP \eqno(4.2)$$
$$PPPA \quad PPAP \quad PAPP \quad APPP $$
$$ PPPP~, $$
and, hence, $1+4+6+4+1=16$ different physical ground states. The periodic
boundary
condition implies integer modding, whereas the anti-periodic one yields only
half-integer modes in the oscillator (mode) expansion of $\j(z)$.

The sector $AAAA$ has only half-integer world-sheet fermionic modes and has
no zero
fermionic modes. It is therefore quite similar to the NS-sector of the
conventional
superstring \cite{gsw}, and comprises just one {\it space-time bosonic} state
$\ket{{\rm phys}}=\ket{0,p}$ with $p^2=0$, which follows as a solution to
the physical
state condition
$$Q\ket{{\rm phys}}=0~,\eqno(4.3)$$
subject to the usual (Siegel's) gauge conditions
$$\r_0\ket{{\rm phys}}=\o_0\ket{{\rm phys}}=0~,\eqno(4.4)$$
in that sector. Each of the four sectors in the second line of eq.~(4.2)
contains only
{\it one} real fermionic zero mode, which obeys the Clifford algebra as a
 consequence
of eqs.~(2.6) and (2.8). It leads to a {\it space-time fermionic} ground
state of the
form $u_{\e}\ket{1/2,p}$, characterized by the momentum $p$ and the spinor
wave function
 $u_{\e}$ with a polarization $\e$. Being the only physical state in this
sector, it
satisfies the on-shell conditions
$$\g^{\ula}p_{\ula}u_{\e}=p^2=0~,\eqno(4.5)$$
which also follow from eq.~(4.3). Hence, those four sectors are all of the
Ramond-type
and represent space-time fermions in the theory.

The fifth line of eq.~(4.2) yields the sector in which all the world-sheet
fermions have integer modes and, in particular, anticommuting integer zero
modes.
The physical
state condition (4.3) should then be supplemented by the gauge conditions
$$\x_0\ket{{\rm phys}}=d^{\bar{a}}\ket{{\rm phys}}=0~,\qquad \bar{a}=\bar{1},
\bar{2}~,
\eqno(4.6)$$
in addition to that of eq.~(4.4). Given a formal vacuum state $\ket{0,p}$ in
the
extended (by ghosts) Fock space for that sector, one can construct more
states at the
same level, when acting on that vacuum by the operators $d_0^a~,\quad a=1,2\,$:
$$\ket{0,p},\qquad d^a_0\ket{0,p}~,\qquad \ve_{ab}d^a_0d^b_0\ket{0,p}~.
\eqno(4.7)$$
The physical states are distinguished by eq.~(4.3) which, being applied to a
linear
combination of all the states in eq.~(4.7), picks up the longitudinal vector
$$\bar{p}_a d_0^a\ket{0,p}~,\qquad p^2=0~,\eqno(4.8)$$
as the only physical solution ({\it cf} \cite{mm2}). This state obviously has
vanishing ghost and excitation numbers and, therefore, represents
one {\it space-time bosonic} physical state in the $PPPP$ sector.

Similarly, six ground states corresponding to the third line of eq.~(4.2)
turn out
to be space-time {\it bosons}, whereas four ground states from the fourth
line of
that equation are all  space-time {\it fermions}.

Since in any $SO(2,2)$-invariant theory, there should be an equal number of
states with positive and negative norms at each ``spin'' level, the six new
bosonic states should
form a space with indefinite norm, the same being true for all space-time
fermions: only a half of them (say, from the second line of eq.~(4.2)) should
have positive norms, while the others (from the fourth line of eq.~(4.2))
should
then be with negative norms.

Putting all together, we thus have 8 bosonic and 8 fermionic states in the
theory,
which is apparently space-time supersymmetric in $2+2$ dimensions. To fix
space-time
representations of all the physical states, we now consider the ground state
in the
$AAAA$ (=NS) sector of the theory and find first its quantum numbers.

The NS sector of the closed $N=2$ superstrings was analyzed by Ooguri and Vafa
\cite{ov1}
in the non-covariant $N=2$ formulation. They noticed the NS ground state to be
a massless ``scalar'', and argued about the absence of other sectors in the
theory. However, the arguments of ref.~\cite{ov1} were based on the possibility
to {\it continuously} interpolate between what they called NS and R sectors by
using ``Wilson line'' operators associated with the Abelian charge in the
$N=2$ superconformal algebra (sect.~2). The sectors we discussed above form a
{\it discrete} set in that sense, but they are related by space-time
supersymmetry
(sect.~5).

To uncover the nature of the NS ``scalar'', the gauge-fixed $N=2$ string
action
$$I_{\rm g-f}=\fracmm{1}{\p}\int d^2 zd^2\q d^2\bar{\q}\,\h_{a\bar{b}}X^a
\bar{X}^{\bar{b}}\equiv \fracmm{1}{\p}\int d^4 Z\, K_0(X,\bar{X})~,
\eqno(4.9)$$
and the vertex operator describing the emission of the NS ground state with
momentum
$k^{\ula}=(k^a,\bar{k}^{\bar{a}})$,
$$V_{\rm c}=\fracmm{\k}{\p}:\exp\[i(k\cdot\bar{X}+\bar{k}\cdot X)\]:~,
\eqno(4.10)$$
written in terms of the $N=2$ chiral scalar superfields $X(Z,\bar{Z})=
X(z,\bar{z},\q,\bar{\q})$, can be used to calculate the $N=2$ string
tree scattering amplitudes and then analyze the effective field theory
\cite{ov1}. The $\k$ is the $N=2$ closed string coupling constant.

In particular, the 3-point tree amplitude takes the form \cite{ov1}
$$A_3\sim\VEV{\left.V_{\rm c}\right|_{\q=0}(0)\cdot\int d^2\q d^2\bar{\q}\,
V_{\rm c}(1)\cdot V_{\rm c}\left|_{\q=0}(\infty)\right.}$$
$$=\k\left(k_1\cdot\bar{k}_2 -\bar{k}_1\cdot k_2\right)^2
\equiv \k c_{12}^2~,\eqno(4.11)$$
where the super-M\"obius invariance had been used to fix three points
on the $N=2$ super-Riemannian sphere. The $A_3$ is non-vanishing  and
apparently non-covariant with respect to the Lorentz group $SO(2,2)$.

The calculation of the 4-point $N=2$ closed string tree amplitude yields
the result \cite{ov1}
$$A_4\sim\int d^2z\,\VEV{\left.V_{\rm c}\right|_{\q=0}(0)\cdot\int d^2\q
d^2\bar{\q}\,V_{\rm c}(z)\cdot\int d^2\q d^2\bar{\q}\,V_{\rm c}(1)\cdot
 V_{\rm c}\left|_{\q=0}(\infty)\right.}$$
$$=\fracmm{\k^2}{16\p}\int d^2z\,\abs{\fracmm{1}{(1-z)^2}t(t+2)+
\fracmm{c_{12}c_{34}}{z}+\fracmm{c_{23}c_{41}}{1-z}}^2\abs{z}^{-s}
\abs{1-z}^{-t}$$
$$=\fracmm{\k^2}{\p}F^2\fracmm{\G(1-s/2)\G(1-t/2)\G(1-u/2)}{\G(s/2)
\G(t/2)\G(u/2)}~,\eqno(4.12)$$
where $s,t,u$ are the Mandelstam variables, $s=-(k_1\cdot\bar{k}_2 +
\bar{k}_1\cdot k_2),~ etc.$, and
$$F\equiv 1- \fracmm{c_{12}c_{34}}{su} - \fracmm{c_{23}c_{41}}{tu}=0~.
\eqno(4.13)$$
The $A_4$ amplitude is vanishing because of the (non-trivial) kinematical
identity (4.13), which has also been proved in ref.~\cite{ov1}. The
vanishing kinematical factor $F$ in eq.~(4.12) is needed for the
consistency of the theory, otherwise there would be massive poles in the
amplitude which are absent in the spectrum \cite{ov1}. It was just the
reason for the general conjecture made by Ooguri and Vafa that all the
trees $A_n$ at $n\geq 4$ should actually vanish \cite{ov1} in the $N=2$
string theory.

The effective field theory which reproduces the above-mentioned $N=2$ string
tree amplitudes in the NS sector had also been constructed
in ref.~\cite{ov1}, and it has turned out to be the Plebanski theory of {\it
self-dual gravity} (SDG) \cite{pl} in $2+2$ dimensions:
$$S_{\rm P} =\int d^4x\,\left(\ha \h^{a\bar{b}}\pa_a\f\pa_{\bar{b}}\f
+\fracmm{\k}{3}\f\pa_a\pa_{\bar{a}}\f\ve^{ab}\ve^{\bar{a}\bar{b}}\pa_b
\pa_{\bar{b}}\f\right)~.\eqno(4.14)$$
The NS ``scalar'' $\f$ should therefore been identified with the deformation
of
the flat K\"ahler potential: $K=K_0 + 2\k\f$. The equations of motion in
the Plebanski theory (4.14) can be rewritten to the form of the SDG equations
for the four-dimensional curvature tensor $R$ to be constructed from the
metric
$g_{a\bar{b}}=\pa_a\pa_{\bar{b}}K$ \cite{pl},
$${}^*R=R~,\eqno(4.15)$$
which are obviously $SO(2,2)$ covariant. Hence, the NS ``scalar'' is not just
an ordinary scalar, but non-trivially transforms under the Lorentz
transformations.
Since the effective field theory has turned out to be the SDG in $2+2$
dimensions,
it now becomes clear that the NS ``scalar'' represents the self-dual
``graviton'',
which has  the non-vanishing helicity  $(+2)$
and ``spin'' $2$ to be defined with respect to the ``little group'' $GL(1)$
and
the Lorentz group $SO(2,2)$, respectively \cite{s3}.

The situation with the $N=2$ open strings is quite similar. The corresponding
tree
string amplitudes  are to be the ``square roots'' of the closed ones, due
to holomorphic factorization \cite{klt}. In addition, the Chan-Paton factors
$\{\L^I\}$ can now be introduced, as usual in open string theory \cite{ci}.
An appropriate $N=2$ open
string world-sheet is the upper-half-superplane $\S$, whose boundary is
$\pa\S\neq 0$. The open $N=2$ string vertex operator takes the same form
(4.10), which is supposed  to be restricted to $\pa\S$.

In particular, the calculation of the 3-point amplitude yields \cite{ms}
$$A_3^{\rm o}\sim\VEV{\left.V_{\rm o}\right|_{\q=0}(0)\cdot\int d^2\q \,
V_{\rm o}(1)\cdot V_{\rm o}\left|_{\q=0}(\infty)\right.}
=\k_{\rm o} c_{12}(-if^{IJK})~,\eqno(4.16)$$
where the open $N=2$ string coupling constant $\k_{\rm o}$ and the Lie algebra
structure constants $f^{IJK}$ have been introduced,
$$f^{IJK}=\tr(\L^I,\[\L^J,\L^K\])~.\eqno(4.17)$$
This 3-point function is again non-vanishing and non-covariant with respect to
the Lorentz group.

Recently, the $N=2$ open-string 4-point tree amplitude has been calculated
\cite{ms}:
$$A^{\rm o}_4\sim\int^1_0 dx\,\VEV{\left.V_{\rm o}\right|_{\q=0}\cdot
\int d^2\q\,
V_{\rm o}(x)\cdot \int d^2\q\, V_{\rm o}(1)\cdot V_{\rm o}\left|_{\q=0}\right.
(\infty)}$$
$$=\fracmm{\k_{\rm o}^2}{4}F\fracmm{\G(1-2s)\G(1-2t)}{\G(2u)}~.\eqno(4.18)$$
It also vanishes since $F=0$ \cite{ms}.

The effective field theory describing the $N=2$ open string trees takes the
form
\footnote{The $c_R$ is the quadratic Casimir operator eigenvalue for the gauge
group generators.}
$$S_{\rm PS} = \fracmm{1}{c_R}\int d^4x\,\left(-\ha \h^{a\bar{b}}\pa_aV
\pa_{\bar{b}}V + \fracmm{\k_{\rm o}}{3}\ve^{\bar{a}\bar{b}}V\pa_{\bar{a}}V
\pa_{\bar{b}}V\right)~,\eqno(4.19)$$
and it is just the self-dual Yang-Mills (SDYM) action of Parkes \cite{p8}.
The
equations of motion in the Parkes theory (4.19) can be rewritten to the
standard
form of the SDYM equations:
$${}^*F=F~,\eqno(4.20)$$
which are explicitly covariant with resepct to the $SO(2,2)$ Lorentz
transformations.
The Parkes ``scalar'' $V$ appears to be a SDYM (non-covariant)
potential with  non-trivial Lorentz
transformation properties. The SD interpretation
of the theory (4.19) suggest to identify this ``scalar''
 with a self-dual vector particle, whose
helicity is $+1$ and ``spin'' 1 \cite{s2}. The SDYM
equations also follow from requring the vanishing of the sigma-model
beta-functions in the $N=2$ heterotic string theory with the Yang-Mills
background \cite{gn1}.

We are now in a position to discuss the full space-time covariant
formulation of
the $N=2$ fermionic strings in terms of the equivalent local quantum
supersymmetric
field theory, when all 16 sectors of the string theory being taken into
account.

\section{Space-Time Symmetries and Covariant Actions}

As we have learned from the previous sections, the effective field theory
describing $N=2$ open string modes (or left-moving modes of the $N=2$ closed
string) in $2+2$ dimensions should be
\begin{itemize}
\item {\it Lorentz covariant} with respect to
$SO(2,2)\cong SL(2,{\bf R})\otimes SL(2,{\bf R})$,
\item space-time {\it supersymmetric},
\item {\it self-dual}.
\end{itemize}
In addition, it comprises just $8_{\rm B}\oplus 8_{\rm F}$ degrees of freedom
in the open case, and $16_{\rm B}\oplus 16_{\rm F}$ degrees of freedom in the
closed case. The space-time supersymmetry actually implies an invariance of
the theory under a larger group $SL(\left. 2\right| N)\otimes
SL(\left. 2\right| N)$, where the number $N$ of space-time supersymmetries is
yet to be determined.

The covariant description of the self-dual supersymmetry and supergravity in
$2+2$ dimensions has recently been developed in refs.~\cite{kng1,gnk3,kng4}.
The existence of {\it Majorana-Weyl} (MW) spinors in $2+2$ dimensions is the
key point in all those constructions. In particular, in the real (Majorana)
representation of the $(2+2)$-dimensional Dirac $4\times 4$ matrices
$\G^{\ula}$,
$$\{ \G^{\ula},\G^{\ulb} \}=2\h^{\ula\ulb}~,\quad \h^{\ula\ulb}={\rm diag}
\,(-,-,+,+)~,\eqno(5.1)$$
the $\G_5$-matrix is {\it real},
$$\G_5=\G^1\G^2\G^3\G^4~,\qquad \G^2_5=1~.\eqno(5.2)$$
The MW spinors transform in the fundamental (real) representation of one of
the $SL(2,{\bf R})$ factors in the Lorentz group, and, hence, they have only
$2/2=1$ degree of freedom on-shell \cite{kng1}. The self-dual vector particle
and MW spinor are naturally united into one $N=1$ supersymmetric self-dual
{\it vector} multiplet with $1_{\rm B}\oplus 1_{\rm F}$ on-shell components.
The self-dual {\it scalar} $N=1$ supermultiplet, comprising a {\it real}
scalar and a MW spinor, can also be constructed \cite{kng1}.

In {\it extended} supersymmetry, there exists $N=2$ self-dual vector multiplet
\cite{gnk3}, comprising
a self-dual vector, two MW spinors of the same chirality and a real scalar
($2_{\rm B}\oplus 2_{\rm F}$ on-shell components). Naively, one could expect
that the $N=4$ SDYM would also follow the same pattern, and thus contain
$4_{\rm B}\oplus 4_{\rm F}$ components in its on-shell spectrum, but it turns
out {\it not} to be the case \cite{kng4}. The $N=4$ SDYM actually needs twice
as many degrees of freedom for its definition even on-shell \cite{kng4}. The
$N=4$ SDYM has the following on-shell field contents \cite{s3}
$$\left(A\du\ula I,G\du{\ula\ulb} I, \r^I,\Tilde\l^I, S\du {\hat i}I, T\du
{\hat i}I\right)~,\eqno(5.3)$$
where $G_{\ula\ulb}$ is anti-symmetric and anti-self-dual, $\r$ and
$\Tilde{\l}$ are anti-MW and MW spinors, respectively, $S_{\hat i}$ and
$T_{\hat i}$ are scalars, $\hat{i}=\hat{1},\hat{2},\hat{3}$; all fields
being real and in the adjoint representation of a gauge group $G$, $I=1,2,
\ldots,\dim G$.

The situation with self-dual supergravities (SDSG's) in $2+2$ dimensions is
quite similar \cite{s2,kng4}. There exist the $N$-extended SDSG's up to the
$N\leq 4$, which comprise $N_{\rm B}\oplus N_{\rm F}$
on-shell degrees of freedom in $2+2$ dimensions, but it is no longer true for
$N>4$ SDSG's, in which the number of the on-shell degrees of freedom is
doubled:  $2N_{\rm B}\oplus 2N_{\rm F}$ \cite{s2}.

There are just $8_{\rm B}\oplus 8_{\rm F}$ on-shell components in the
irreducible $N=4$ SDYM multiplet, and just $16_{\rm B}\oplus 16_{\rm F}$
on-shell componets in the irreducible multiplet of the $N=8$ SDSG. But in
order to investigate the possibility to identify them with the $N=2(4)$ open
and closed string modes respectively, as was suggested by Siegel \cite{s2,s3},
one should still compare the interactions to be defined separately, in terms
of the $N=2$ open and closed string diagrams and in terms of the $N=4$ SDYM
or $N=8$ SDSG Feynman graphs, respectively. First, however, one needs
covariant
actions for the supersymmetric self-dual field theories.

As for the {\it ordinary} SDYM theory, its covariant action has recently
been
proposed by Kalitzin and Sokatchev \cite{ks5} in the {\it harmonic} space,
which
can be $M_{2+2}\otimes S^2$. Their action reads \footnote{The $\pm$ or the
number indices (in parentheses) mean the $U(1)$ charge.}
$$S_{\rm KS} = \int d^4x d^2u\,\tr\left[\L^{(-3)\a}\pa^+_{\a}\left( e^{\cv}
D^{++}e^{-\cv}\right)\right]~.\eqno(5.4)$$
The harmonic coordinates on the sphere $S^2=SU(2)/U(1)$ are introduced as
$$\left(\begin{array}{c} u^{+\a'} \\
u^{-\a'}\end{array}\right)\in SU(2)~,\qquad \ve_{\a'\b'}u^{+\a'}u^{-\b'}=1~.
\eqno(5.5)$$
The relevant derivatives in eq.~(5.4) are
$$D^{++}= u^{+\a'}{\pa\over \pa u^{-\a'}}~,\quad
\pa^+_{\a}=u^{+\b'}\pa_{\a\b'}~,\quad \ula=(\a\a')~.\eqno(5.6)$$
The Yang-Mills theory in the harmonic space is defined by the relations
$$\pa_{\a\b'}\to \cd_{\a\b'}\equiv \pa_{\a\b'} + A_{\a\b'}(x)~,$$
$$\cd^{\pm}_{\a}\equiv u^{\pm\b'}\cd_{\a\b'}~,\to D^{++}A^+_{\a}(x,u)=0~,
\eqno(5.7)$$
which imply the constraint \cite{ks5}
$$ \[ D^{++},\cd^+_{\a}\]=0~.\eqno(5.8)$$
The SDYM condition $F_{\a'\b'}=0$ can now be rewritten to the form \cite{ks5}
$$\[\cd^+_{\a},\cd^+_{\b}\]=0~,\eqno(5.9)$$
because of the relation $F_{\a\b}^{++} = \ve_{\a\b}u^{+\a'}u^{+\b'}
F_{\a'\b'}$. The zero-curvature condition (5.9) can be explicitly solved:
$$ \cd^+_{\a}(x,u)=e^{-\cv(x,u)}\pa^+_{\a}e^{\cv(x,u)}~,\eqno(5.10)$$
and then eq.~(5.8) becomes the equation of motion for the harmonic field
$\cv(x,u)$. The action which reproduces that equation is just the
Kalitzin-Sokatchev action (5.4). This action does give the SDYM condition
on-shell, when varying with respect to the Lagrange field $\L^{(-3)}(x,u)$.
On the other hand, the Lagrange multiplier itself turns out {\it not} to be
a propagating field but a pure gauge in the KS action (5.4), which is
invariant under the gauge transformations \cite{ks5}:
$$\L^{(-3)}(x,u)\to \L^{(-3)}(x,u) +\pa^{+\a}b^{(-4)}~.\eqno(5.11)$$

The substitution $\cv\to V^{++}=e^{\cv}D^{++}e^{-\cv}$, which is
non-local in the harmonic space, but {\it local} in space-time, makes the
action (5.4) to be the action of a {\it free} theory. Hence, there is
{\it no}
scattering in the quantized KS theory, which was demonstrated to a great
extent in
ref.~\cite{moy}. The $N=1$ and $N=2$ supersymmetric extensions of the KS
action
can also be constructed along the similar lines \cite{do}, and there should
be
no principal problems to write down the KS-type actions for the $N$-extended
SDSG's of ref.~\cite{kng4} for $N\leq 4$. Nevetheless, they are all actually
the
{\it free} theories. The lesson we should learn  is that the ``purely''
self-dual covarinat field theories. i.e. those without any additional
propagating degrees of freedom, are actually free and do not have scattering.

The $N=4$ SDYM theory has a quite different covariant component action, when
the
``extra'' fields needed to complete the supermultiplet are playing the role of
the Lagrange multipliers for the rest of the fields and vice versa
\cite{s3,kng4},
$$\Lag{\,}_{\rm SP}^{N=4~{\rm SDYM}}=\, - \fracm 12 G^{\ula\ulb\,I}
(F\du{\ula\ulb}I- \half\e\du{\ula\ulb}{\ulc\uld} F\du{\ulc\uld} I) + \fracm 12
(\nabla_\ula S\du{\hat i} I)^2 - \fracm 12 (\nabla_\ula T\du{\hat i}I)^2$$
$$ + i (\r^I \s^\ula D_\ula \Tilde\l^I)
 - i f^{I J K} \left[ (\Tilde \l^I \a_{\hat i} \Tilde\l^J)
S\du{\hat i} K + (\Tilde\l^I \b_{\hat i} \Tilde\l^J) T\du{\hat i} K
\right] ~,\eqno(5.12)$$
where $\nabla_\ula$ and $D_\ula$ are the gauge-covariant derivatives, $\s^\ula$
represent the Dirac matrices in the 2-component notation for the $SO(2,2)$
Lorentz group,
$$\G^{\ula} = \left(\begin{array}{cc} 0 & \s^{\ula} \\ \tilde{\s}^{\ula} & 0
\end{array}\right)~,\eqno(5.13) $$
in which the $\G_5$ is supposed to be diagonal. The $\a$ and $\b$ matrices are
the second independent set of the gamma matrices for the $SO(4)$ or $SO(2,2)$
internal symmetry \cite{kng4}. The similar action does also exist for the
$N=8$ SDSG \cite{s2}.

It is not known how to rewrite that actions  in the maximally extended
superspace. However, one can construct the SP-type self-dual field theory
actions at lower $N$, when introducing the Lagrange multipliers to the fields
of the $N$-extended SDYM and SDSG \cite{kng4}. For instance, the $N=1$
PS-type SDYM action in the $N=1$ superspace takes the form  \cite{kng4}
$$S{\,}_{\rm SP}^{N=1~{\rm SDYM}} = \int d^4xd^2\theta\,\L^{\a\,I} W\du\a I ~~,
\eqno(5.14) $$
where the chiral real $N=1$ SDYM superfield strength $W^I_{\a}$ and the
Lagrange superfield $\L^{\a\,I}$ have been introduced. In components,
eq.~(5.14)
reads
$$\eqalign{S{\,}_{\rm SP}^{N=1~{\rm SDYM}}=
\int d^4 x\Big[ & -\half G^{\ula\ulb\, I}
(F_{\ula\ulb} {}^I (A) - \half \e\du{\ula\ulb}{\ulc\uld}
F_{\ulc\uld} {}^I (A) )
+i\r ^{\a\, I} (\s^\ula)_{\a\Dot\b} \nabla_\ula\Tilde\l^{\Dot \b\, I}
\Big] ~~, \cr }\eqno(5.15) $$
where all the auxiliaries have been eliminated, and the following propagating
fields have been introduced:
$$\eqalign{
 &\L_\a {}^I | = \r \du\a I ~~, ~~~~ \nabla_\a\L_\b {}^I | =
(\s^{\ula\ulb})_{\a\b} G_{\ula\ulb} {}^I + \ldots~, \cr
 & W_\a {}^I | = \l_\a {}^I ~~, ~~~~ \nabla_\a W_\b {}^I | =
(\s^{\ula\ulb})_{\a\b} F_{\ula\ulb} {}^I (A) + \ldots~. \cr }\eqno(5.16) $$

The action (5.14) can be combined with a similar action for the three
self-dual
{\it scalar} $N=1$ supermultiplets \cite{kng4} to be accompanied by the
corresponding three scalar Lagrange multiplier $N=1$ superfields. It will
result
in the $N=1$ superspace representation of the $N=4$ SDYM theory, which is
quite
convenient for analyzing quantum loops. The $N=8$ SDSG in terms of the $N=1$
self-dual superfields and their $N=1$ Lagrange multipliers follows the same
pattern.

The SP-type actions (5.12) and (5.14) are covariant and have non-vanishing
 3-point vertices. Hence, there is a non-trivial scattering of covariant
objects in that theories. In addition, their quantum loops are all vanishing,
because of the non-renoramlization theorem in (extended) supersymmetry
\cite{ggrs}. It actually implies the $N$-extended {\it superconformal}
invariance of those theories \cite{s2,s3}:
$$\def\normalbaselines{\baselineskip20pt
\lineskip3pt \lineskiplimit3pt }
\def\mapright#1{\smash{
\mathop{\longrightarrow}\limits^{#1}}}
\def\mapdown#1{\Big\downarrow
\rlap{$\vcenter{\hbox{$\scriptstyle#1$}}$}}
\matrix{
SO(2,2) & \cong & SL(2)\otimes SL(2) & \mapright{\rm susy} & SL(2|N)\otimes
SL(2|N) \cr
&& \mapdown{\rm conf} && \mapdown{\rm s-conf} \cr
SO(3,3) & \cong & SL(4) & \mapright{\rm susy} & SL(4|N)\cr}\eqno(5.17)$$

Since the $N=2(4)$ superstring has no massive modes and has to be equivalent
to the effective local quantum field theory, there should be an exact
correspondence between the string and field amplitudes. The space-time
covariance,
supersymmetry and self-duality apparently force the equivalent quantum field
theory
to be the $N=4$ SDYM in the open case, and the $N=8$ SDSG in  the closed case,
{\it provided} the string physical states are all belong to the one {\it
irreducible} supermultiplet. Given the covariant equivalence, it implies the
need to re-examine the status of the known $N=2$ superstring amplitudes (see
the previous section) with respect to their space-time symmetries, if any.

\section{Covariance in Loops Forces Them to Vanish}

The $N=2(4)$ fermionic string theory has the same dilemma as the conventional
superstrings: it has to have reducible gauge symmetry generators in order to
be explicitly space-time covariant and supersymmetric, but quantization can
only be
efficiently performed in a non-covariant gauge. That's why our strategy is to
consider the quantized $N=2$ strings and impose covariance and supersymmetry
as the consistency conditions.

In the Polyakov approach to the $N=2$ string theory, the partition function
takes the form
$$Z =\int \[ Dg D\c DA DxD\j\] e^{-I_{\rm BS}}~,\eqno(6.1)$$
where the action  $I_{\rm BS}$ has been introduced in eq.~(2.4). A topology
of a closed $N=2$ super-Riemannian surface is characterized by the two
integers: the genus $g$ and the first Chern class $c$. In the component
approach we adopted, the genus $g$ is simply related to the Euler
characteristics $\c(\S)$ of the Riemann surface,
$$\c(\S)\equiv \fracmm{1}{2\p}\int_{\S} d^2z\,\sqrt{g}R=2-2g~,\eqno(6.2)$$
whereas the first Chern class $c$ can be identified with the ``instanton''
number for the Abelian gauge field Wilson lines along the cycles of the
surface.

The deformations of the world-sheet metric can be orthogonally decomposed
as usual \cite{mm2,dp}
$$\d g_{mn}=\{\d\s g_{mn}\}\oplus \{P_1\d v\}_{mn}\oplus Ker\,P^{\dg}_1~,
\eqno(6.3)$$
with respect to the natural norm
$$ ||\d g_{mn}||^2=\int_{\S} d^2z\,\sqrt{g}g^{lm}g^{np}\d g_{ln}\d g_{mp}~,
\eqno(6.4)$$
where $(P_1\d v)_{mn}\equiv \de_m(\d v)_n + \de_n(\d v)_m -g_{mn}\de^p
(\d v)_p$; $\d v_p$ and $\d\s$ are the infinitesimal parameters of
reparametrizations and Weyl transformations, respectively. Similarly,
one has \cite{mm2}
$$\d\c_{n}=\{\g_n\d\L\}\oplus \{P_{1/2}\d\z\}\oplus Ker\,P^{\dg}_{1/2}~,
\eqno(6.5)$$
where $(P_{1/2}\d\z)_n\equiv\tilde{\de}_n\d\z-\ha\g_n\g^m\tilde{\de}_m\d\z$,
$\tilde{\de}_n\equiv \de_n -iA_n$; $\d\z$ and $\d\L$ are the infinitesimal
superconformal and super-Weyl transformation parameters, respectively. The
deformations of the Abalian gauge field follow the same pattern \cite{mm2}:
$$\d A_n =(P_0\d\ve)_n\oplus(\hat{P}_0\d\q)_n\oplus \cap\,Ker\,(P_0^{\dg},
\hat{P}_0^{\dg})~,\eqno(6.6)$$
where $(P_0\d\ve)_n=\pa_n\d\ve$, $(\hat{P}_0\d\q)_n=
\e_{nm}\pa_m\q$; $\d\ve$ and $\d\q$ are the infinitesimal phase and chiral
gauge transformation parameters, respectively. The spaces of moduli
deformations $Ker\,P_1^{\dg}$,
$Ker\,P^{\dg}_{1/2}$ and $\cap Ker\,(P_0^{\dg},\hat{P}_0^{\dg})$,
as well as the
conformal Killing (CK) spaces $Ker\,P_1$, $Ker\,P_{1/2}$ and
$\cap Ker\,(P_0,\hat{P}_0)$, are all finite-dimensional,
their dimensions being
 determined in general by $g$ and $c$, in accordance to
the Riemann-Roch theorem. In particular, for a surface with $g\geq 2$, one
has $(3g-3)$ complex moduli for the metric, $2(2g-2)$ complex fermion moduli
for the two ``gravitini'', and $g$ complex moduli for the Abelian vector gauge
 field. For the torus $(g=1)$, the CK vectors (CKV's), spinors (CKSp's) and
scalars  (CKS's) are to
be taken into account \cite{mm2}. For our purposes, it is enough to notice
that the partition function can be reduced to a finite-dimensional integral
over the $N=2$ supermoduli space $\cm_{g,c}$ at the given genus $g$ and the
first Chern class $c$, which schematically takes the form
$$Z=\sum_{g,c=0}^{\infty}\sum_{\rm spin\atop structures}
\int_{\cm_{g,c}}d({\rm WP})
\fracmm{ \{ {\rm zero-mode~contributions}\} }{Vol({\rm CKV})Vol({\rm CKSp})
Vol({\rm CKS})}\prod_{\{\a\}}{\det}'\D_{\{\a\}}~,\eqno(6.7)$$
where $d({\rm WP})$ denotes the Weyl-Petersson measure,
$\prod_{\{ \a\} }\det'\D_{\{ \a\} }$ symbolically represents
the contributions of all
non-zero modes, and the zero modes are treated separately. The appearance of
the Weyl-Petersson measure has to be expected if one insists on the
modular invariance of the theory, whereas all the non-zero mode contributions
should actually cancel altogether, as a consequence of boson-fermion
supersymmetry. It can also be understood in the world-sheet terms because of
the natural pairing between the coordinates and ghosts in the $N=2$ string
theory, which was also responsible for the actual absence of the non-zero
physical modes. The detailed one-loop calculations of ref.~\cite{mm2} also
support that conjecture.

Being transported along closed cycles, the world-sheet fermions can either be
periodic or anti-periodic, and all that information is just equivalent to
assigning a spin structure \cite{dp}. The partition function is supposed to be
 defined as
 a sum over all spin structures. The integration over Wilson lines of the
Abelian gauge field of the $N=2$ string changes spin structures, and it is thus
equivalent to the summation over all {\it continuous} changes of
fermionic boundary conditions. However, it does not cover the discontinuous
changes in eq.~(4.2), since it would then contradict the Lorentz invariance,
in particular. It was always assumed that all sectors of the $N=2$ string
should contribute with the same phase to the partition function, and that
argument was always supported by the fact that all the contributions from
various sectors are separately modular invariant (see e.g., \cite{mm2}).
Without any symmetry restrictions, it was indeed a good assumption to reduce
an arbitrariness in the thoory, since any other reasoning like
factorizability or unitarity are rather doubtful in $2+2$ dimensions. But
when the $SO(2,2)$ Lorentz symmetry and space-time supersymmetry have to
be taken into account, the situation becomes quite different. It is
space-time supersymmetry which dictates the relative phases between really
different (discontinuous) sectors of the $N=2$ string theory, and which forces
separate equal contributions to sum to zero in the total partition function.

The self-dual space-time supersymmetry actually forces all string
amplitudes to vanish, at any positive genus and ``instanton'' number. There is
no need even for the maximal number of supersymmmetries: just one space-time
supersymmetry in self-dual and space-time covariant theory makes the job done.
As was calculated in ref.~\cite{bgi}, the one-loop 3-point amplitude in the
NS-type sector of the $N=2$ string theory is non-vanishing and severely
IR-divergent. However, the contribution of space-time fermions from the
R-type sectors of the string theory is just supposed to cancel it.

{}From this viewpoint, the status of the non-vanishing 3-point tree string
amplitudes discussed in sect.~4 should be reconsidered. The tree amplitudes
can be defined inside each sector independently of the existence of any other
sectors in the theory. Therefore, those tree amplitudes cannot be ``made to
vanish''. However, they actually describe the scattering of the non-covariant
quantities. Applying the Lorentz boost to any of the 3-point tree
amplitudes of sect.~4 results in the multiplicative redefinition
of the coupling constant as the only change \cite{p8}. Therefore, there is
no Lorentz-invariant scattering to be described by those amplitudes. It
clearly matches with the often repeated statement about the
``no tree scattering in SDYM'' \cite{w}. In other
words, there is no covariant amplitude which would reduce, say, to  eq.~(4.11)
in an appropriate gauge. If we now take a look on the equivalent quantum field
theory of eq.~(5.12), we immediately notice only the presence of the 3-point
vertices in the Lagrangian, which relate {\it different} sectors of the string
theory. Though there is no covariant scattering inside each of the sectors in
eq.~(4.2), it is consistent with space-time covariance and supersymmetry to
have 3-point interactions between SD fields and their Lagrange multipliers.
Since the states corresponding to the Lagrange multipliers are supposed to
have
negative or zero norms, this result is not very optimistic for the future
prospects of the theory under consideration.

\section{Conclusion}

The outcome of our discussion can be summarized as follows.

The critical $N=2$ fermionic strings live in 2 complex dimensions, the Lorentz
group being $SO(2,2)$ and the signature being $(-,-,+,+,)$. They have only
zero
modes as the physical states. The open $N=2$ fermionic string spectrum is
space-time supersymmetric and comprises 8 bosonic and 8 fermionic states. The
$N=2$ closed string spectrum has 16 bosonic and 16 fermionic states. The
spin contents of that supermultiplets exactly coincides with that of the
$N=4$ supersymmetric SDYM and the $N=8$ SDSG, respectively, which are
the equivalent local quantum space-time supersymmetric field theories for the
$N=2$ fermionic strings. Those results were first outlined by Siegel
\cite{s1,s2,s3} on the ground of
pure symmetry considerations. We have seen above how his ideas can be supported
inside the NSR-like formulations for the $N=2$ and $N=4$ fermionic strings.
However, our results do not give the ultimate proof of an actual
presence of the {\it maximal} space-time supersymmetry: it is the case under
the additional assumption about the irreducibility of the zero mode
supermultiplets. The existence of the GS-type formulation of the $N=2(4)$
strings,
recently proposed by Siegel \cite{s4}, with the fermionic degrees of freedom
to be
represented by the $SO(2,2)$ {\it spinors}, gives the alternative covariant
and
space-time supersymmetric description of the same theory but it does not rule
 out
the possibility to have the space-time supersymmetry which is not maximal.

The space-time covariance and supersymmetry in the theory of $N=2$ fermionic
strings
imply the absence of scattering for any covariant quantities inside of each
sector of
that string theory. The only allowed type of covariant scattering is the
3-point
scattering between the self-dual fields and their Lagrange multipliers
corresponding
to the different sectors, when the specific vertices being given by the
appropriate Siegel-Parkes-type action. In those supersymmetric field theory
PS-type
actions, the Lagrange multipliers are necessarily the propagating fields.
All the
string loop amplitudes vanish as the consequence of the {\it alternative}
 sum over spin structures, which is the only consistent with the space-time
covariance and supersymmetry. Therefore, the $N=2$ fermionic string theory is
equivalent to the supersymmetric local quantum field theory which looks like a
topological field theory.

The huge degeneracy of the $N=2(4)$ non-compact fermionic string theory is
ultimately
responsible for the mutual cancellation of various non-trivial contributions
to its tree and loop amplitudes. Therefore, one should expect that even slight
deformation of this theory can make it to be very non-trivial and the
amplitudes
to be non-vanishing. One of the way has recently been suggested in
 ref.~\cite{lpwx} by using the background charge to be represented by the
additional term in the stress tensor along the standard lines of the
Dotsenko-Fateev construction \cite{df} in conformal field theory, or,
equivalently, at the expense of the non-trivial dilaton expectation value
in the sigma-model approach to strings. Those modified $N=2$ fermionic
strings turn out to live in $1 + 2m$ dimensions, $m=1,2,\ldots$, and do not
have vanishing 4-point amplitudes \cite{bl}. Therefore, the theory of the
extended fermionic strings still has a potential to be non-trivial not only
topologically but also geometrically, i.e. with non-vanishing tree and string
loop amplitudes, all having a rich geometrical structure.

\section{Acknowledgements}

The author thanks H.~Nicolai, O.~Lechtenfeld and C.~Preitschopf for useful
discussions.

\end{document}